\pdfoutput=1
\documentclass[11pt]{article}

\usepackage[]{KINSPEAK}

\usepackage{times}
\usepackage{latexsym}

\usepackage[T1]{fontenc}
\usepackage[utf8]{inputenc}

\usepackage{microtype}

\usepackage{inconsolata}

\usepackage{float}

\usepackage{subcaption,graphicx}
\graphicspath{ {./img/} }

\usepackage{amsmath}
\usepackage{mathtools}
\usepackage{amssymb}
\usepackage{outline}
\usepackage{booktabs}

\usepackage{bm}      % for \bm macro

\newcommand{\figref}[1]{Figure~\ref{fig:#1}}
\newcommand{\tblref}[1]{Table~\ref{table:#1}}

\newcommand{\B}[1]{\textbf{#1}}
\newcommand{\U}[1]{\underline{#1}}

\title{KinSPEAK: Improving speech recognition for Kinyarwanda via semi-supervised learning methods}

\author{Antoine Nzeyimana \\
  University of Massachusetts, Amherst, USA \\
  \texttt{anzeyimana@umass.edu} \\}

\begin{document}
\maketitle

\begin{abstract}
Despite recent availability of large transcribed Kinyarwanda speech data, achieving robust speech recognition for Kinyarwanda is still challenging. In this work, we show that using self-supervised pre-training, following a simple curriculum schedule during fine-tuning and using semi-supervised learning to leverage large unlabelled speech data significantly improve speech recognition performance for Kinyarwanda. Our approach focuses on using public domain data only. A new studio-quality speech dataset is collected from a public website, then used to train a clean baseline model. The clean baseline model is then used to rank examples from a more diverse and noisy public dataset, defining a simple curriculum training schedule. Finally, we apply semi-supervised learning to label and learn from large unlabelled data in five successive generations. Our final model achieves 3.2\% word error rate (WER) on the new dataset and 15.6\% WER on Mozilla Common Voice benchmark, which is state-of-the-art to the best of our knowledge. Our experiments also indicate that using syllabic rather than character-based tokenization results in better speech recognition performance for Kinyarwanda.
\end{abstract}

\section{Introduction}

Expanding access to automatic speech recognition (ASR) technology to the large number of speakers of low resource languages can improve the quality of their interactions with computing and multi-media devices. Moreover, it makes it possible to perform analytics and retrieval on the large amount of speech data produced by these communities. This is more significant for speakers of low resource languages such as Kinyarwanda and other language communities that disproportionately use more oral than written forms of communication.

\begin{table}[ht!]
\centering
\caption{Results on Mozilla Common Voice benchmark}
\resizebox{0.98 \columnwidth}{!}{
\begin{tabular}{l c c}
\toprule
\B{Methods}  &  \B{CER \%}  &  \B{WER \%} \\
\midrule
Conformer + CTC  &  8.4  &  25.3 \\
+ Self-PT  &  6.8  &  21.2 \\
+ Curriculum  &  6.4  &  20.1 \\
+ Semi-SL Gen. 1  &  5.7  &  18.0 \\
% +Semi-SL Gen 2  &  5.3  &  16.9 \\
% +Semi-SL Gen 3  &  5.0  &  16.5 \\
% + Semi-SL Gen. 4  &  \B{\U{4.8}}  &  \B{\U{15.9}} \\
+ Semi-SL Gen. 5  &  \B{\U{4.7}}  &  \B{\U{15.6}} \\
\midrule
NVIDIA NeMo: CTC  &  5.5  &  18.4 \\
NVIDIA NeMo: Transducer  &  5.7  &  16.3 \\
\bottomrule
\end{tabular}
}
\label{table:results_highlight}
\vspace{-.18in}
\end{table}

However, achieving robust speech recognition for a language such as Kinyarwanda is challenging. First, the lack of tone markings in the written form of a tonal language requires a lot more diverse data for speech recognition to work well in practice. Second, a morphologically rich language (MRL) like Kinyarwanda has a very large vocabulary, making it difficult to achieve low word error rates (WER) on public benchmarks. For example, while the data scarcity problem for Kinyarwanda has been largely alleviated, thanks to Mozilla Common Voice~\cite{ardila2020common} (MCV) crowd-sourcing efforts, off-the-shelf models accuracy on the MCV test set still has room for improvement. These challenges motivate us to explore language-specific treatments towards more robust speech recognition for Kinyarwanda.

Recent advances in deep learning techniques for end-to-end speech recognition and the availability of open source frameworks and datasets allow us to empirically explore different ways to improve ASR performance for Kinyarwanda. While recent experimental reports and studies~\cite{ravanelli2021speechbrain,ritchie2022large} have shown improvement in ASR for Kinyarwanda, mostly via self-supervised pre-training (Self-PT) representations such as wav2vec2.0~\cite{baevski2020wav2vec}, there haven't been exploration of using Kinyarwanda-only speech data for Self-PT pre-training and how to improve performance beyond using Self-PT representations. In this work, we report empirical experiments showing how ASR performance for Kinyarwanda can be improved though Self-PT pre-training on Kinyarwanda-only speech data, following a simple curriculum learning schedule during fine-tuning and using semi-supervised learning (Semi-SL) to leverage large unlabelled data. We also compare the impact of two different tokenization techniques on ASR performance on this particular language.

Our approach focuses on leveraging available public domain data. First, we collect 22,000 hours of Kinyarwanda-only speech data from YouTube and use it for Self-PT pre-training. Then, we fine-tune a clean baseline model on studio-quality transcribed readings from JW.ORG website\footnote{\label{jw.org}\url{https://www.jw.org/rw/isomero/}}. The clean baseline model is then used to rank MCV training examples by character error rates (CER), so that we can formulate a curriculum learning schedule for a final model trained on the combined dataset. Since JW.ORG website readings are long and not segmented, we developed an easy to use mobile application that annotators used to align audio data to text, yielding about 89 hours of clean, segmented and transcribed speech data. A simple curriculum learning schedule was then devised to train an intermediate model in 6 stages, by subsequently introducing harder and harder examples to the training process. Additionally, we experiment with semi-supervised learning (Semi-SL) by transcribing and filtering audio segments from our YouTube speech data, and then adding this new dataset to our original training dataset and resuming training on the new combined dataset. This process is repeated for five generations. Our final model achieves 15.6\% WER on MCV benchmark version 12 and 3.2\% WER on the new dataset from JW.ORG website.

Specifically, we make the following contributions:
\begin{itemize}
    \item We empirically show how ASR performance for Kinyarwanda can be improved by leveraging available public domain speech data only, and combining three different technique, namely, self-supervised pre-training, curriculum-based fine-tuning and semi-supervised learning.
    \item We devise a simple mobile design methodology for speech-text alignment to enable collection of labelled utterances from long unaligned speech and text data. This design allows crowd-workers to easily annotate data using mobile devices, making the whole process scalable.
    \item By exploring two alternative tokenization techniques, we empirically show that using syllabic instead of the more common character-based tokenization leads to better performance in end-to-end ASR for Kinyarwanda.
\end{itemize}

\section{Related work}

The primary transcribed speech dataset we used comes from the Mozilla Common Voice (MCV) speech corpus~\cite{ardila2020common}. This dataset contains more than 2,000 hours of transcribed Kinyarwanda speech which was collected through a crowd-sourcing effort led by a Kigali-based company called Digital Umuganda~\footnote{\url{https://digitalumuganda.com}}.

The secondary public domain speech data we used was downloaded from Jehovah's Witnesses website, JW.ORG~\textsuperscript{\ref{jw.org}}. Due to the high-quality multi-lingual speech and text data on JW.ORG, the data from this website has been subject to studies in language technology research, most notably machine translation~\cite{agic2019jw300}. However, due to license restrictions, this data cannot be redistributed outside of the JW.ORG website. To the best of our knowledge, we are the first ones to experiment speech recognition using JW.ORG data.

As the MCV dataset evolved through multiple versions, there have been several studies and experimental results~\cite{ritchie2022large,ravanelli2021speechbrain,kuchaiev2019nemo} reported on the dataset. The best results are generally obtained by fine-tuning pre-trained models such as wav2.vec2.0~\cite{baevski2020wav2vec} which are typically pre-trained on large English-only or multi-lingual speech data. Our approach focuses on using a moderately sized Kinyarwanda-only speech data downloaded from YouTube.

Our model architecture is inspired by the work in~\cite{zhang2020pushing} which is based on recent end-to-end speech recognition architectures and techniques including Convolution-augmented transformers (Conformer)~\cite{gulati2020conformer}, SpecAugment~\cite{park2020specaugment}, self-supervised pre-training~\cite{baevski2020wav2vec,hsu2021hubert} and connectionist temporal classification (CTC)~\cite{graves2006connectionist}.

There have been a number of studies of using syllable-based tokenization for ASR in different languages~\cite{savithri2023suitability,zhou2018comparison}. However its effectiveness varies from language to language. To the best of our knowledge, we are the first ones to investigate the suitability of syllable-based tokenization for Kinyarwanda ASR.

\section{Methods}

\subsection{Collecting studio-quality transcribed utterances via speech-text alignment}

Curriculum learning has been shown to improve robustness in automatic speech recognition~\cite{braun2017curriculum}. Given that the publicly available MCV dataset is diverse and noisy in many cases, defining a measure for ranking the examples is hard. While one can use the number of up-votes and down-votes provided by MCV corpus~\cite{ardila2020common}, these may not be reliable given that most validated examples have two up-votes and there can only be a maximum of three votes per example. Instead, we chose to use a extrinsic assessment by training a clean baseline model using a studio-quality dataset and then using the clean baseline model to rank MCV examples based on character error rates (CER).

To develop the clean baseline model, we relied on publicly available readings from JW.ORG~\textsuperscript{\ref{jw.org}} website. The readings are made of a web-page with a media player playing the content of the page. However, the readings are typically long and the text is not aligned with the utterances in most cases. Therefore, it is not trivial to segment the data for ASR training and there is no automated tool that can reliably do the alignment.

\begin{figure}[!ht]
 \centering
     \setlength{\fboxsep}{0pt}%
    \setlength{\fboxrule}{1pt}%
    \fbox{\includegraphics[scale=0.2]{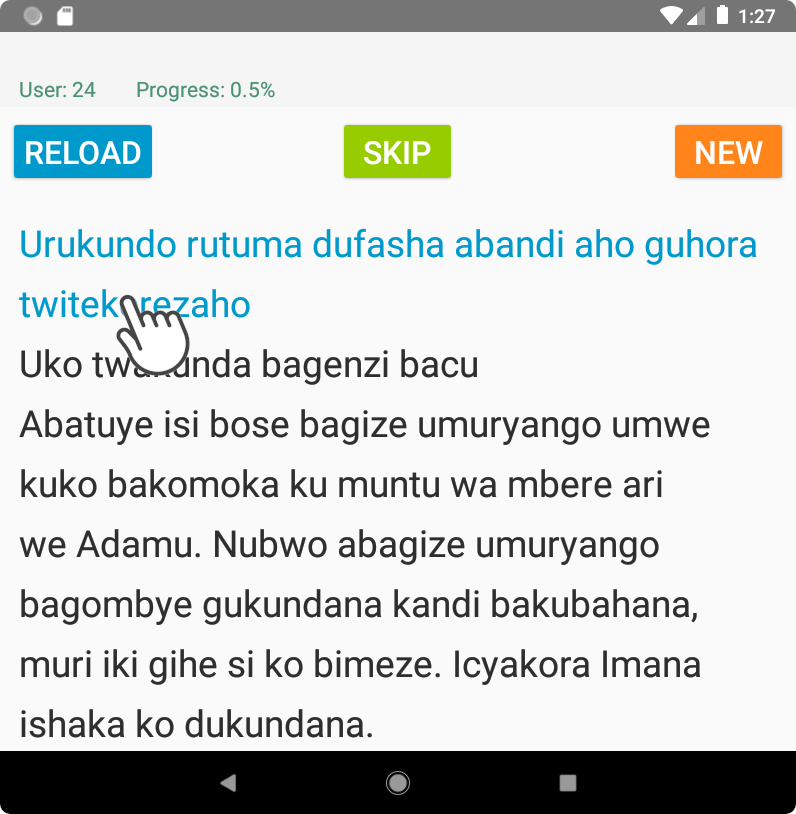}}
 \captionof{figure}{\label{fig:huza_imvugo} Speech-text alignment mobile interface. The annotator is asked to touch the last word played by the audio clip before the pause. The selected segment is highlighted and then cut out and the process repeats until the end of the document.}
 \vspace{-.2in}
\end{figure}

In order to align speech and text data from JW.ORG website~\textsuperscript{\ref{jw.org}}, we developed an easy to use mobile application and asked volunteers to align the text to speech segments using the developed application. This required first downloading the HTML pages with the associated audio clips, extracting clean text from HTML and marking long-enough silences in the audio clips as segmentation boundaries. The clean text, the link to the audio clip and silence markings (time index in the audio clip) were then stored as metadata for the mobile application. 

During its use, the application would play each audio clip while showing the text and pausing at each silence point for the user to mark the last spoken word in the displayed text. Five Kinyarwanda native speakers then volunteered to annotate the metadata. These annotators were asked to listen to the audio clips and mark the silence points by touching the last spoken word on the touch-screen displaying the text. After the touch, the text segment up to the last touched word is then highlighted and then cut out, and the process of listening and touching continues until the end of the audio clip or the end of the text. The index of the last spoken word in the text and the corresponding silence index in the audio clip were then stored on a back-end server and later post-processed for creating the final dataset for ASR model training. A screen capture of the mobile interface is shown in~\figref{huza_imvugo}.

\subsection{ASR Model architecture}

Automatic speech recognition (ASR) aims at mapping a sequence of speech representations ${X = \{\mathbf{x}_i \in \mathbb{R}^d\}}$ to a sequence of text units ${Y = \{y_j \in \mathcal{V}\}}$ in a given language. The modeling problem is then to find the most probably sequence of text units $\hat{Y}$ among all vocabulary combinations $\mathcal{V}*$ ("strings") given a speech input, i.e.: 
\begin{align}\label{eq:main_eq}
    \hat{Y} &= \underset{Y\in \mathcal{V}*}{\arg\max} \;p(Y\mid X).
\end{align}
The current practice in large vocabulary speech recognition is to use end-to-end neural networks for joint optimization instead of designing independent discrete components. One approach treats ASR as a sequence prediction problem (i.e. sequence-to-sequence) and thus uses encoder-decoder architectures~\cite{chan2016listen} like machine translation or text summarization. A different approach uses only an encoder to learn a monotonic mapping of encoder outputs to target text units. This is motivated by the fact that speech is mostly produced at constant speed, where each text unit can be mapped to a specific temporal segment in the input speech. In this case, a loss function based on Connectionist Temporal Classification (CTC)~\cite{graves2006connectionist} can be used to optimize the encoder outputs over all valid token alignments.

\begin{figure}[!ht]
 \centering
    \includegraphics[scale=0.4]{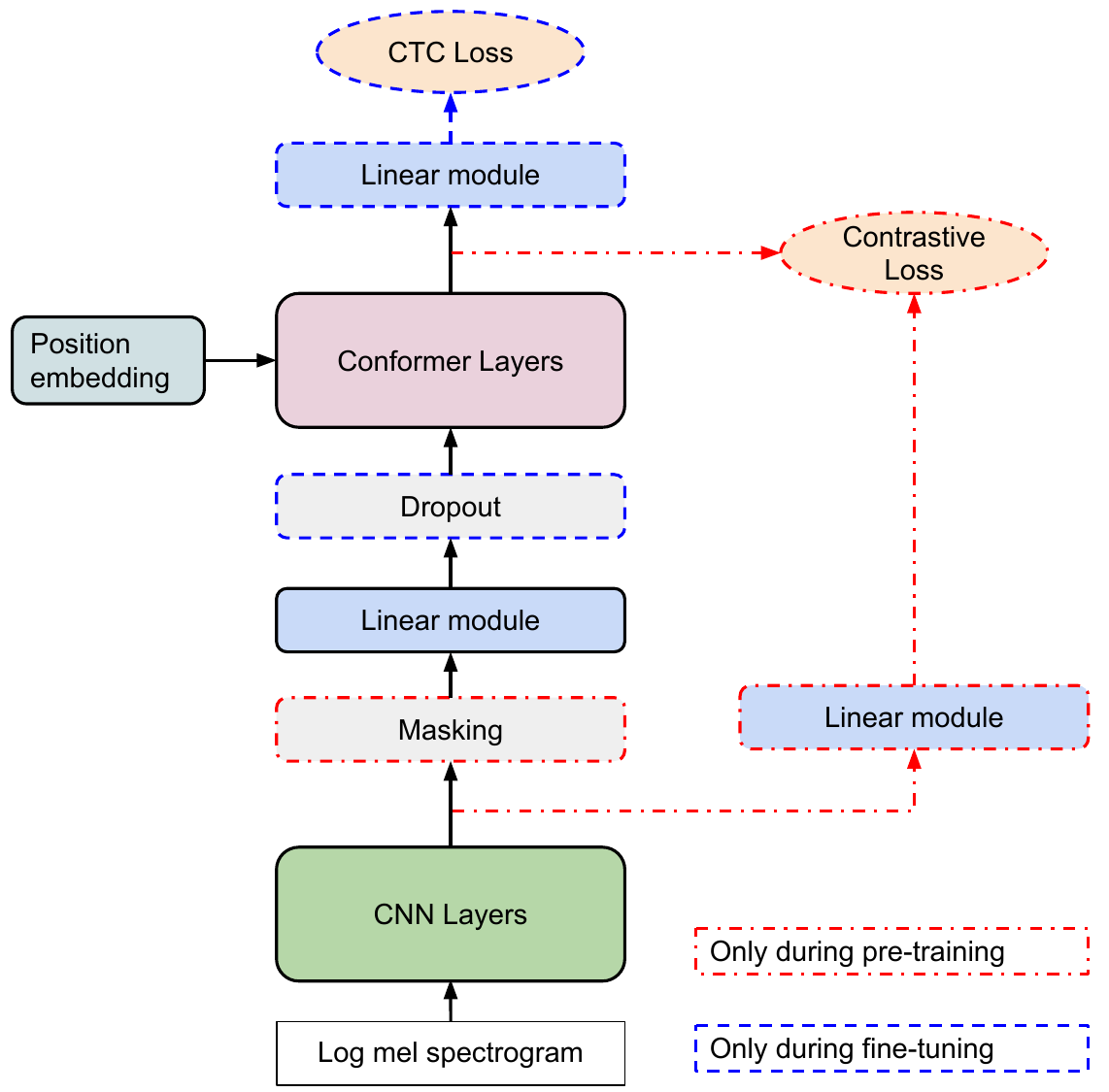}
 \captionof{figure}{\label{fig:architecture} ASR Model architecture}
 \vspace{-.1in}
\end{figure}

Our ASR model architecture is shown in~\figref{architecture}. The model is based on the conformer architecture~\cite{gulati2020conformer} and uses a contrastive objective for self-supervised pre-training in a way closely similar to~\cite{zhang2020pushing}. The input to the model consists of the log mel-spectrogram of an utterance. The input is then processed by a convolutional sub-sampling module which creates a reduced-length input to the rest of the network.

During the pre-training phase, the output of the convolutional sub-sampling module is masked and sent to the conformer layers on one hand, and also projected through a linear layer and then used for contrastive learning. The masking parameters come from Wav2Vec2.0~\cite{baevski2020wav2vec}. However, differently from Wav2Vec2.0, we do not use quantization before our contrastive loss calculation. Instead, we use the linear projection layer before contrasting masked positions. This method was proposed by~\cite{zhang2020pushing} and is sufficient for pre-training our network. During normal training, which we call fine-tuning, both the masking and contrastive projection are removed and a new projection layer is added on top of the conformer layers for Connectionist Temporal Classification (CTC)~\cite{graves2006connectionist} -based training.

Differently from~\cite{zhang2020pushing}, our conformer model formulation uses untied position embedding method called TUPE~\cite{kerethinking} which adds position embedding vectors as a bias to the multi-head self-attention sub-module. We also use a pre-LayerNorm configuration~\cite{nguyen2019transformers} of the conformer modules.

\subsection{Multi-staged curriculum schedule for training}

Curriculum learning~\cite{bengio2009curriculum} has been shown to improve robustness in speech recognition~\cite{braun2017curriculum}. We designed a multi-stage curriculum schedule, whereby the model learns from cleaner examples earlier than noisy examples. In our formulation, we follow a coarse-grained multi-stage schedule. At stage 0, we train the model on JW.ORG-only training set which contains about 80 hours of data. At each subsequent stage, we double the amount of training examples by adding harder and harder examples from MCV training set, until the whole MCV training set is covered. We train each intermediate stage for 10 epochs, while the final stage containing all training data is trained for 49 epochs. At each new stage, we keep the model weights, while resetting the optimizer state and learning rate schedule.

After completing the training on JW.ORG and MCV dataset, we explore the use of semi-supervised learning (Semi-SL) by transcribing utterances from the YouTube dataset and adding those examples to the training set and resuming training. This method was also explored in~\cite{zhang2020pushing}, where it's called noisy student training (NST). Our approach follows a curriculum schedule, whereby easier examples are added first, based on ranking by an external language model. Specifically, we use our model trained on JW.ORG and MCV datasets to transcribe a large portion of YouTube utterances, then rank their transcripts and finally pick the top ranked examples, add them to the training set and then resume training. For ranking transcribed YouTube utterances, we use both the CTC beam search~\cite{zenkel2017comparison} decode scores and log-probabilities produced by an external tokenization-aligned language model. We apply thresholds to both scores to pick the top ranked utterances and their transcriptions. Unlike the NST scheme in~\cite{zhang2020pushing}, we don't restart training from the pre-trained model at each generation. Instead, we resume training from the model trained on JW.ORG and MCV data. In our experiments, we repeated this semi-supervised learning process for four generations to get our final model.

\begin{table}[!ht]
\caption{Syllable-based tokenization: Kinyarwanda vowels, consonants and consonant clusters.}
    \centering
\resizebox{0.98 \columnwidth}{!}{
\renewcommand{\arraystretch}{1.2}
\begin{tabular}{|c c c c c|}
\hline
i  &  u  &  o  &  a  &  e \\
\hline
b  &  c  &  d  &  f  &  g \\
h  &  j  &  k  &  m  &  n \\
p  &  r  &  l  &  s  &  t \\
v  &  y  &  w  &  z  &   \\
\hline
bw  &  by  &  cw  &  cy  &  dw \\
fw  &  gw  &  hw  &  kw  &  jw \\
jy  &  ny  &  mw  &  my  &  nw \\
pw  &  py  &  rw  &  ry  &  sw \\
sy  &  tw  &  ty  &  vw  &  vy \\
zw  &  pf  &  ts  &  sh  &  shy \\
mp  &  mb  &  mf  &  mv  &  nc \\
nj  &  nk  &  ng  &  nt  &  nd \\
ns  &  nz  &  nny  &  nyw  &  byw \\
ryw  &  shw  &  tsw  &  pfy  &  mbw \\
mby  &  mfw  &  mpw  &  mpy  &  mvw \\
mvy  &  myw  &  ncw  &  ncy  &  nsh \\
ndw  &  ndy  &  njw  &  njy  &  nkw \\
ngw  &  nsw  &  nsy  &  ntw  &  nty \\
nzw  &  shyw  &  mbyw  &  mvyw  &  nshy \\
nshw  &  nshyw  &  njyw  &    & \\
\hline
\end{tabular}
    % }
}
\label{table:tokens}
 \vspace{-.15in}
\end{table}

\subsection{Syllable-based tokenization}

Many end-to-end deep learning models for speech recognition use character-based tokenization. However, like many other Bantu languages, Kinyarwanda only has open syllables~\cite{walli1986representation}, whereby each syllable ends with a vowel. Also, when Kinyarwanda orthography is being taught in elementary school~\footnote{\url{https://elearning.reb.rw/course/view.php?id=293}}, students are first introduced to vowels, then simple consonants and finally consonant clusters as basic orthographic units. Therefore, we wanted to explore this syllable-based tokenization, where output tokens are either vowels, simple consonants and consonant clusters as thought in elementary school Kinyarwanda orthography.

In our experiments, we empirically compare the effectiveness of this syllable-based tokenization against the more common character-based tokenization. \tblref{tokens} shows the basic text units of Kinyarwanda orthography which we use in our experiments. In addition to these basic orthographic units, we add foreign characters `x' and `q' which are typically used in foreign words and names. Therefore, the main difference between character- and syllable-based tokenization is that syllable-based tokenization allows consonant clusters in the vocabulary while character-based tokenization doesn't (i.e. it must learn to produce consonant clusters from single consonants during inference). We also add six basic punctuation marks as part of our vocabulary, namely, full stops (.), commas (,), question marks (?), exclamation marks (!), colons (:), and apostrophes ('). This allows our models to produce basic punctuation from speech without without the need for a dedicated punctuation restoration model~\cite{tilk2015lstm}. During evaluation, we omit these punctuation marks to have fair comparison with other open source models which don't include punctuation marks.

\section{Experimental setup}

\begin{table*}[!ht]
\caption{Ablation results: Character error rates (CER \%) and word error rates (WER \%) on the validation (Dev.) and test sets across different training configurations. Best ablation results are shown in bold. JW: JW.ORG dataset. MCV: Mozilla Common Voice dataset.}
    \centering
\resizebox{0.98 \textwidth}{!}{
% {\renewcommand{\arraystretch}{1.05}% for the vertical padding
\begin{tabular}{l c c c c c c c c c c}
\toprule
         \multicolumn{3}{c}{~} & \multicolumn{2}{c}{\B{JW Dev.}} & \multicolumn{2}{c}{\B{JW Test}} & \multicolumn{2}{c}{\B{MCV Dev.}} & \multicolumn{2}{c}{\B{MCV Test}} \\
\B{Tokenization}  &  \B{Self-PT}  &  \B{Curriculum}  &  \B{CER}  &  \B{WER}  &  \B{CER}  &  \B{WER}  &  \B{CER}  &  \B{WER}  &  \B{CER}  &  \B{WER} \\
\midrule
Character  &  -  &  -  &  3.8  &  14.4  &  3.8  &  14.2  &  6.0  &  20.6  &  8.5  &  26.0 \\
Syllable  &  -  &  -  &  3.5  &  13.1  &  3.4  &  12.8  &  5.7  &  19.8  &  8.4  &  25.3 \\
\midrule
Character  &  Yes  &  -  &  1.5  &  4.9  &  1.4  &  4.7  &  5.1  &  17.8  &  6.9  &  22.2 \\
Syllable  &  Yes  &  -  &  1.4  &  4.6  &  1.3  &  4.4  &  5.0  &  17.3  &  6.8  &  21.2 \\
\midrule
Character  &  Yes  &  Yes  &  \B{1.3}  &  4.2  &  \B{1.1}  &  3.9  &  4.8  &  16.8  &  6.7  &  20.8 \\
Syllable  &  Yes  &  Yes  &  \B{1.3}  &  \B{3.9}  &  \B{1.1}  &  \B{3.7}  &  \B{4.7}  &  \B{16.3}  &  \B{6.4}  &  \B{20.1} \\
\bottomrule
    \end{tabular}
    % }
 }
\label{table:ablation_results}
 \vspace{-.15in}
\end{table*}

\subsection{JW.ORG speech data gathering and text-speech alignment}

The text and audio clips from the JW.ORG website were identified and collected through web crawling. This process happened in August 2021 and we identified 792 documents with 139 hours of speech data. The text was extracted from HTML documents using jsoup~\footnote{\url{https://jsoup.org/}} Java library. The mobile application for text to speech alignment was developed for the Android operating system. Any user of the application was assigned to annotate all documents in a random order. While more than 50 anonymous users attempted to volunteer to participate in the annotation process, only data from five participants were used to compile the final dataset. These included one author of this paper, one trained and paid annotator and two volunteers who are known and related to the author and another anonymous volunteer. Because of the tedious work involved, only the paid annotator annotated all documents while others annotated about 5\% of all documents. The other annotators' data was used to evaluate the inter-annotator agreement. The inter-annotator agreement ratio is calculated as the number of agreeing silence marker-last spoken word pairs divided by the total number of silence markers in the commonly annotated documents. This inter-annotator agreement ratio was 90.7\% between the author and the paid annotator and 82\% between the author and the other annotators. Most of the disagreements happened in handling text in parentheses and biblical references. In the post-processing stage, we removed data segments that were deemed too short ({audio length $<$ 2 seconds} or {text length $<$ 5 characters}) or too long ({audio length $>$ 30 seconds} or {text length $>$ 400 characters}). We also removed those segments whose number of syllables per second was more than 1.3 standard deviations from the average. Finally, we got a final clean audio-text dataset with total 86 hours of speech, which we randomly split into training, validation and test sets in the ratios of 90\%, 5\% and 5\% respectively.

\subsection{ASR model implementation}

Our ASR model was implemented with PyTorch~\cite{paszke2019pytorch} framework version 1.13.1. The CNN component is made of two layers of 3x3 convolutions with stride of 2. The conformer component is made of 16 layers of conformer blocks, each with 768 hidden dimension, 8 attention heads and 3072 feed-forward dimension. The whole model has about 229 million parameters. The input to the model are log mel-spectrograms which are computed with 1024-point short-time Fourier transform (STFT) on 25 milliseconds of raw audio frames (sampled at 16 Hz), with a hop length of 10 milliseconds, and using 80 mel filters.

\subsection{Training process}

We collected about 22,000 hours of unlabelled YouTube data from 37 channels that publish speech content in Kinyarwanda on various topics including news and political discussions and social conversations. The data was randomly segmented into 5 to 25 seconds-long segments. During contrastive pre-training, we used a mask probability of 0.5 and a static mask length of 10, in a way similar to Wav2Vec2.0~\cite{baevski2020wav2vec}. We used a global batch size of 1.6 hours of data, a peak learning rate of 5e-4, 544,000 training steps, 32,000 warm-up steps and a linear learning rate decay. Pre-training took 25 days on two NVIDIA RTX 4090 GPUs. During ASR model fine-tuning, we used a global batch size of 800 seconds, peak learning rate of 2e-4, 5000 warm-up steps, inverse square root learning rate decay, and the training was done for about 50 epochs in each case. ASR model fine-tuning experiments were done on NVIDIA RTX 3090 GPUs, taking 1.25 seconds per step one one GPU. For decoding, we use CTC beam search~\cite{zenkel2017comparison} algorithm which we implemented in C++. Through experimentation on the development set, we set the beam width to 24.

\begin{table*}[!ht]
\caption{Semi-supervised learning results: Character error rates (CER \%) and word error rates (WER \%) on the validation (Dev.) and test sets through iterative generations. Overall best results are shown in bold and underlined. JW: JW.ORG dataset. MCV: Mozilla Common Voice dataset}
    \centering
\resizebox{0.98 \textwidth}{!}{
% {\renewcommand{\arraystretch}{1.05}% for the vertical padding
\begin{tabular}{c c c c c c c c c c}
\toprule
         \multicolumn{1}{c}{~} & \B{Unlabelled data} & \multicolumn{2}{c}{\B{JW Dev.}} & \multicolumn{2}{c}{\B{JW Test}} & \multicolumn{2}{c}{\B{MCV Dev.}} & \multicolumn{2}{c}{\B{MCV Test}} \\
\B{Generation}  &  \B{size (hours)}  &  \B{CER}  &  \B{WER}  &  \B{CER}  &  \B{WER}  &  \B{CER}  &  \B{WER}  &  \B{CER}  &  \B{WER} \\
\midrule
1  &  1K   &  1.2      &  3.4      &  1.1      &  3.4      &  4.4      &  15.3      &  5.7      &  18.0 \\
2  &  3K   &  \B{\U{1.1}}  &  3.2      &  \B{\U{1.0}}  &  3.2      &  4.2      &  14.8      &  5.3      &  16.9 \\
3  &  10K  &  \B{\U{1.1}}  &  \B{\U{3.1}}  &  \B{\U{1.0}}  &  \B{\U{3.1}}  &  4.1      &  14.6      &  5.0      &  16.5 \\
4  &  10K  &  1.2      &  3.3      &  \B{\U{1.0}}  &  3.2      &  4.0  &  14.3  &  4.8  &  15.9 \\
5  &  20K  &  1.2      &  3.2      &  \B{\U{1.0}}  &  3.2      &  \B{\U{3.9}}  &  14.2  &  \B{\U{4.7}}  &  \B{\U{15.6}} \\
\midrule
\multicolumn{2}{l}{NVIDIA Nemo: CTC} &  2.3  &  9.7  &  2.2  &  9.4  &  4.3  &  15.3  &  5.5  &  18.4 \\
\multicolumn{2}{l}{NVIDIA Nemo: Transducer} &  2.2  &  8.7  &  2.2  &  8.8  &  4.5  &  \B{\U{14.1}}  &  5.7  &  16.3 \\
\bottomrule
    \end{tabular}
    % }
 }
\label{table:final_results}
 \vspace{-.15in}
\end{table*}

\subsection{Evaluation}

We use both character error rates (CER \%) and word error rates (WER \%) as evaluation metrics. CER are particularly important for this case because Kinyarwanda is a morphologically-rich language and thus tends to have an open vocabulary. Furthermore, as shown in~\ref{error_analysis}, there exist some writing ambiguities where a Kinyarwanda reader may consider multiple spellings of the same word as legit. Therefore, using WER metrics alone is not sufficient. We use TorchMetrics package~\cite{detlefsen2022torchmetrics} version 0.11 for our metrics computation.

%
% 2. ASR Model details:
% Implementation framework and versions
% Model dimensions and size: CNN details, conformer details, total number of parameters
% Data pre-processing details: segmentation lengths, resampling configuration (to 16 Hz), log mel spectrogram configuration,
% Pre-training hyper-parameters
% Fine-tuning hyper-parameters
% Multi-stage configuration: number of stages, MCV data size for each stage
% Learning rate schedule, batch size, peak learning rate, optimizer, weight decay, 
% Pre-training infrastructure and run time
% Fine-tuning infrastructure and run time
% Inference: CTC beam search configuration, Lexicon size, C++, shared library
% Inference run time: speedup
% Semi-SL generations and configuration: Can LM NLL < 1.5, 1 gen: first 1000 hours (33% yield), 2nd gen: 3000 hours (51% yield), limited by inference speeds
%

%
% 3. Language models
% Only used for ranking CTC beam search outputs
% Language models configuration & size: char-/syllable-based, n M parameters, Transformer-based
% Language model training details: infrastructure, training time
% Language model training corpus: 2.4 GB of monolingual Kinyarwanda text from books and online newspapers and blogs
% 

% 4. 

\section{Results and discussion}

\subsection{Effects of tokenization, self-supervised pre-training and curriculum learning}

Our first ablation results are presented in~\tblref{ablation_results}. We compare character-based tokenization and syllable-based tokenization across three training and evaluation setups: with and without self-supervised pre-training, and finally adding the devised curriculum-based training schedule. These results were obtained after about 50 epochs of training on JW.ORG and MCV datasets. Overall, syllable-based tokenization consistently achieves lower error rates than character-based tokenization. Using the curriculum learning schedule also produces significantly better results. Without self-supervised pre-training (Self-PT) , error rates of the baseline models are increased significantly. We also remark that model performance on JW.ORG dataset (JW) is much better than on MCV data; this is because JW.ORG data has very little environmental noise.

\subsection{Semi-supervised learning results}

Our best and final model resulted from using semi-supervised learning (Semi-SL) to leverage the unlabelled YouTube dataset. The results are presented in~\tblref{final_results}. We applied Semi-SL in five generations, first using top 1,000 hours of YouTube dataset, then 3,000 hours, 10,000 hours twice and finally 20,000 hours of data. This best model achieved 1.0\% CER / 3.2\% WER on JW.ORG test set and 4.7\% CER / 15.6\% WER on MCV test set.

Our final model performance is compared to two open source ASR models for Kinyarwanda available on Hugging Face website, which were trained the MCV dataset of similar size to ours. There are two versions of models with NVIDIA NeMo~\cite{kuchaiev2019nemo}, one based on conformer architecture with CTC-based training\footnote{\url{https://huggingface.co/nvidia/stt_rw_conformer_ctc_large}}, the other using encoder-decoder architecture (i.e. Transducer)\footnote{\url{https://huggingface.co/nvidia/stt_rw_conformer_transducer_large}}. While the transducer model achieves WERs comparable to our final model performance, it significantly achieves worse CERs across both JW.ORG and MCV benchmarks.

\begin{table}[!ht]
\caption{Model performance across demographic groups}
    \centering
\resizebox{0.98 \columnwidth}{!}{
% {\renewcommand{\arraystretch}{1.1}% for the vertical padding
\begin{tabular}{c c c c c c}
\toprule
        ~ & \B{\# Train} & \multicolumn{2}{c}{\B{MCV Dev.}} & \multicolumn{2}{c}{\B{MCV Test}} \\
        \B{Group} & (x1000) & CER & WER & CER & WER \\ \midrule
            male   & 517 & 5.1  &  16.5  &  3.8  &  14.0 \\ 
            female & 379 & \B{3.4}  &  \B{12.7}  &  \B{3.2}  &  \B{13.2} \\ 
        \midrule
            teens    & 182 & \B{3.2}  &  \B{13.5}  &  4.6  &  15.1 \\ 
            twenties & 605 & 3.7  &  \B{13.5}  &  \B{3.5}  &  \B{13.5} \\ 
            thirties & 135 & 7.1  &  19.5  &  4.2  &  15.2 \\ 
\bottomrule
    \end{tabular}
    % }
}
\label{table:results_groups}
 \vspace{-.15in}
\end{table}

Since the MCV dataset includes speaker gender and age group labels, we also evaluated our best model across gender and age groups (See \tblref{results_groups}). Overall, examples from female speakers and those from speakers in their twenties resulted in significantly lower WER. Having listened to many noisy MCV utterances, we hypothesise that these performance differences might be more indicative of occupational differences across demographic groups. This is because the MCV Kinyarwanda dataset was mainly contributed to by volunteers throughout their daily activities.

\subsection{Error analysis} \label{error_analysis}

While the achieved WER for Kinyarwanda ASR look particularly higher than those accustomed to in state-of-the-art ASR for high resource languages like English, there are multiple factors that explain this observation. First, Kinyarwanda is morphologically rich and thus has a very large (almost open) vocabulary. But, we also qualitatively found three major types of ambiguities that mostly contribute to the observed high WER. These ambiguities are inherent to written Kinyarwanda and are less dependent on the acoustic model:
\begin{enumerate}
    \item \B{Spelling ambiguities for some consonants}:
    In this case, multiple spellings can be considered legit and understood by readers. Examples:
    \begin{itemize}
        \item[-] \textit{mu{\textcolor{red}{jy}}e / mu{\textcolor{blue}{g}}e} (\textit{`you should'})
        \item[-] \textit{po{\textcolor{red}{r}}itiki / po{\textcolor{blue}{l}}itiki} (\textit{`politics'})
        \item[-] \textit{i{\textcolor{red}{nc}}uti / i{\textcolor{blue}{nsh}}uti} (\textit{`friend'})
    \end{itemize}
    \item \B{Vowel assimilation}:
    This is a known fact~\cite{mpiranya2005sibilant} that vowels at the end of word followed by another word starting with a vowel can be assimilated into the next vowel. Example:
    \begin{itemize}
        \item[-] \textit{avug{\textcolor{red}{a}} \U{a}bantu / avug{\textcolor{blue}{e}} \U{a}bantu} (\textit{`say people ...'}
    \end{itemize}
    \item \B{Loanword and foreign word spelling}:
    There exist no standard for spelling foreign words and names in Kinyarwanda, and different adaptations can be found in writing. Examples:
    \begin{itemize}
        \item[-] \textit{{\textcolor{red}{Patirisiya}} / {\textcolor{blue}{Patricia}}
        \item[-] {\textcolor{red}{Venezuwera}} / {\textcolor{blue}{Venezuela}}}
    \end{itemize}
\end{enumerate}

Therefore, since these ambiguities are common in writing (including benchmark data), they can affect the realized WER.

\section{Conclusion and future work}

This work demonstrates the effectiveness of combining self-supervised pre-training, curriculum learning, and semi-supervised learning methods to achieve better speech recognition performance for Kinyarwanda language. It is also shown that using syllabic tokenization improves upon the more common character-based tokenization for end-to-end ASR for Kinyarwanda. This work can be used by practitioners wanting to develop state-of-the-art ASR systems for other languages that are less prevalent in current language technology research. Future research will focus on incorporating the developed ASR models into language technology applications such as machine translation and information retrieval. We will also investigate on techniques to make the models accessible to mobile architectures such as smartphones, where they can benefit the large community of Kinyarwanda speakers.

\bibliography{kinspeak}
\bibliographystyle{acl_natbib}

% \appendix

% \section{Example Appendix}
% \label{sec:appendix}

% This is a section in the appendix.

\end{document}